\documentclass[12pt]{article}

\usepackage[margin=2.5cm]{geometry}
\usepackage{setspace}
\usepackage{amsmath}
\usepackage{amssymb}
\usepackage{booktabs}
\usepackage{threeparttable}
\usepackage{natbib}
\usepackage{hyperref}
\usepackage{url}
\usepackage{xurl}
\usepackage{graphicx}
\usepackage{caption}
\usepackage{microtype}
\usepackage{siunitx}
\usepackage{array}
\usepackage[T1]{fontenc}
\usepackage[utf8]{inputenc}
\usepackage{rotating} 
\usepackage{longtable} 

\title{Digital Maturity and Technical Efficiency in NHS Acute Trusts:\\
Cross-Sectional Evidence from England}

\author{
Ari Ercole\textsuperscript{1,2,3}\\[0.5em]
\small \textsuperscript{1}Cambridge Centre for AI in Medicine, University of Cambridge, UK\\
\small \textsuperscript{2}Magdalene College, University of Cambridge, UK\\
\small \textsuperscript{3}Cambridge University Hospitals NHS Foundation Trust,\\
\small Addenbrooke's Hospital, Hills Road, Cambridge, UK\\[0.5em]
\small ae105@cam.ac.uk\\
\small ORCID: 0000-0001-8350-8093
}

\date{}

\begin{document}

\maketitle
\thispagestyle{empty}

\begin{abstract}
\noindent Whether investment in digital health technology is associated with differences in hospital productivity is a question of substantial policy relevance, yet interpretation is constrained by challenges in causal identification and prior evidence is mixed. Technical efficiency in NHS acute hospital trusts in England is estimated using Bayesian stochastic frontier analysis. A four-input Cobb--Douglas production function incorporating clinical full-time equivalents, administrative full-time equivalents, non-labour expenditure, and physical capital derived from audited NHS accounts is fitted to 111 acute non-specialist trusts in 2024/25. Digital maturity, measured by the NHS Digital Maturity Assessment, is included in a trust-specific inefficiency equation alongside population deprivation, teaching status, and financial position controls.

The composite digital maturity score is estimated to be negatively associated with technical inefficiency (\(\hat{\gamma} = -0.612\), 95\% credible interval \([-1.289, +0.005]\), \(P(\gamma < 0) = 0.974\)). Trusts in the highest digital maturity quartile are estimated to operate at 98.0\% of their production frontier compared with 93.2\% for the lowest quartile. This gap corresponds to approximately \pounds 20 million of additional cost-weighted activity per trust at mean output levels, or \pounds 1.1 billion in aggregate. Estimates are robust to functional form but are sensitive to the most conservative prior specification. Pillar-level analysis suggests that population health management and care pathway optimisation domains exhibit stronger associations with efficiency than other domains. Catchment deprivation is not estimated to have an independent association with efficiency after controlling for digital maturity.

\medskip\noindent\textbf{Keywords:} stochastic frontier analysis, digital health, NHS, hospital efficiency, Bayesian inference

\medskip\noindent\textbf{JEL codes:} I10, I18, O33, C11
\end{abstract}

\setcounter{page}{1}

\section{Introduction}

NHS acute hospitals face sustained pressure to improve productivity within constrained budgets. Capital investment in digital health technology is frequently proposed as a route to efficiency gains, and NHS England's Digital Transformation programme commits substantial resources on this premise \citep{nhse2021digital}. Whether such investment translates into measurable productivity improvements at the trust level remains, however, poorly quantified.

The theoretical mechanisms through which digital maturity might improve efficiency are well established. Digitally integrated clinical workflows reduce duplicated effort; electronic prescribing reduces medication errors and associated costs; real-time patient flow data supports better bed management; and population health analytics enable preventive pathways that reduce emergency demand. Technology adoption is nonetheless costly, implementation disrupts existing workflows, and  benefits may accrue unevenly across organisational contexts.

Prior evidence on hospital digital maturity and efficiency is modest and methodologically heterogeneous. Studies from the United States report positive associations between health information technology adoption and financial performance and care quality outcomes \citep{bardhan2013electronic}, but the UK context differs substantially in funding mechanisms, incentive structures, and the nature of digital investment. NHS-specific evidence is largely limited to descriptive analyses and aggregate productivity indices that do not control for input usage \citep{nhsi2019tech}.

In this paper, the association between digital maturity and technical efficiency is estimated for all NHS acute non-specialist trusts in England in 2024/25 using Bayesian stochastic frontier analysis (SFA). The Battese-Coelli one-stage approach \citep{bc1995} is adopted to estimate the efficiency equation simultaneously with the production function, avoiding the two-stage bias that arises when efficiency scores are regressed on covariates post-estimation. The model is estimated in a Bayesian framework, allowing prior information and full posterior uncertainty quantification for all parameters.

Three contributions are made. First, fully audited financial data from NHS Trust Annual Accounts are used to construct production function inputs, avoiding the cost-approximation assumptions of prior NHS efficiency studies. Second, physical capital is explicitly included as a production function input and its orthogonality to digital maturity is demonstrated empirically, directly addressing a potential confound. Third, a hierarchical Bayesian model is employed for pillar-level digital maturity effects, enabling differential returns across the seven Digital Maturity Assessment domains to be estimated with appropriate uncertainty quantification.

\section{Data and Methods}

\subsection{Sample}

The analytical sample comprises 111 acute non-specialist NHS trusts in England in 2024/25. Two categories of trust are excluded. Acute specialist trusts (16  organisations) are excluded because their output mix is not comparable to general acute hospitals; such trusts focus on single clinical domains and their cost-weighted activity cannot be placed on a common frontier with general acute providers. Acute multi-service trusts (7 organisations) are excluded because they provide integrated acute and community care services under a single organisational structure. Their total operating expenditure, as recorded in NHS Trust Annual Accounts, therefore encompasses community nursing, social care liaison, and other non-acute activities, while the CWA captures only acute hospital activity. This creates a systematic input-output mismatch: inputs are inflated by community service costs that generate no measured output, producing spuriously low efficiency estimates that reflect the accounting structure rather than productive performance. A ratio of total operating expenditure to cost-weighted activity of up to 2.1 is observed in this group, compared with approximately 1.0--1.1 for standard acute trusts. Both exclusions are defined a priori on grounds of comparability rather than on observed results. The remaining sample covers all acute large, medium, small, and teaching trust subtypes as classified by the NHS England Oversight Framework \citep{nhse2025oversight}.

\subsection{Output}

The output measure is MFF-adjusted cost-weighted activity (CWA) from the NHS National Cost Collection 2024/25 \citep{nhse2025ncc}. CWA aggregates healthcare resource group activity weighted by national reference costs and adjusted for the Market Forces Factor to account for input price variation across geographies. It is a standard output measure in NHS productivity analysis \citep{castelli2011}.

\subsection{Inputs}

Four inputs are included in the production function. Clinical full-time equivalents (FTE) comprises medical staff plus professionally qualified clinical non-medical staff, sourced from NHS Electronic Staff Records (ESR) September 2024 \citep{nhse2025esr}. Administrative FTE comprises support to clinical staff and NHS infrastructure support from the same source. Non-labour expenditure (NLE) is computed as total operating expenditure minus actual staff costs, with both components derived from NHS Trust Annual Accounts (TAC) 2024/25 \citep{nhse2025tac}, ensuring that a common accounting framework is used for both numerator and denominator. Physical capital is the sum of property, plant and equipment and right-of-use assets from TAC balance sheet data. All inputs are log-transformed and centred at their sample means; centring improves sampler geometry without affecting the elasticity interpretation of coefficients.

\subsection{Digital Maturity}

Digital maturity is measured by the NHS Digital Maturity Assessment (DMA) 2025 \citep{nhse2025dma}, a structured self-assessment covering seven domains (`pillars'): Well-Led, Smart Foundations, Safe Practice, Support Workforce, Empower People, Improve Care, and Healthy Populations. Domain scores are averaged to form a composite score, which is standardised to zero mean and unit variance before inclusion in the model. It is acknowledged that the DMA is a survey instrument and may not fully capture operational digital capability. In particular, organisations with long-standing clinical system investment may be underscored relative to externally validated measures such as the HIMSS Electronic Medical Record Adoption Model, potentially attenuating estimated efficiency effects toward zero. A previous Digital Maturity Assessment was conducted in 2023/24. The 2024/25 assessment is used here as it incorporated revised question sets addressing limitations identified in the earlier iteration, providing a more refined measure of organisational digital capability. Results from the 2023/24 assessment are not used.

A related limitation concerns temporal ordering. The cross-sectional design uses contemporaneous digital maturity and efficiency measures, which is appropriate if digital capability and operational efficiency are jointly determined in a stable organisational equilibrium, but precludes assessment of whether efficiency gains lag digital investment. An earlier Digital Maturity Assessment was conducted in 2017; however, the instrument underwent substantial revision between 2017 and 2024/25, with changes to question sets, pillar definitions, and scoring methodology that render the two assessments non-comparable. Using 2017 scores as a lagged exposure for 2024/25 efficiency would therefore introduce measurement error from instrument change, likely exceeding any methodological benefit from temporal separation. Identification of lagged effects awaits a sufficiently long time series of comparable assessments.

\subsection{Controls}

The inefficiency equation includes the Index of Multiple Deprivation score for each trust's catchment population \citep{nhse2026imd}, sourced from NHS acute hospital trust catchment population tables (2024, all admissions), a binary indicator for teaching trust status, and a binary indicator for financial deficit status, both from the NHS Oversight Framework \citep{nhse2025oversight}.

In 2024/25, 89 of 111 trusts (80.2\%) were recorded as being in financial deficit, reflecting the broader NHS financial position in that year. The limited variation  in this indicator reduces its power to identify an independent deficit effect in the inefficiency equation.

\subsection{Production Function}

A Cobb-Douglas production function is specified:

\begin{equation}
\ln(\text{CWA}_i) = \alpha + \beta_1 \widetilde{\ln x}_{c,i} +
\beta_2 \widetilde{\ln x}_{a,i} + \beta_3 \widetilde{\ln x}_{n,i} +
\beta_4 \widetilde{\ln x}_{k,i} + v_i - u_i
\label{eq:prod}
\end{equation}

\noindent where $\widetilde{\ln x}$ denotes mean-centred log inputs for clinical FTE ($c$), administrative FTE ($a$), non-labour expenditure ($n$), and capital ($k$); $v_i \sim \mathcal{N}(0, \sigma_v^2)$ is a symmetric noise term; and $u_i \geq 0$ is technical inefficiency, so that the trust-specific efficiency score is $\text{TE}_i = \exp(-u_i) \in (0,\, 1]$.

Informative priors are placed on the production function coefficients, with prior means set to TAC-derived cost shares: $\beta_{\text{clinical}} \sim \mathcal{N}(0.378,\, 0.15^2)$, $\beta_{\text{admin}} \sim \mathcal{N}(0.197,\, 0.15^2)$, $\beta_{\text{NLE}} \sim \mathcal{N}(0.376,\, 0.15^2)$, and $\beta_{\text{capital}} \sim \mathcal{N}(0.049,\, 0.10^2)$. Cost shares are computed as the ratio of each input's cost to total expenditure using TAC data averaged across the analytical sample; these shares sum to unity by construction. Prior standard deviations of 0.15 (0.10 for capital, which has a smaller share) are selected to be permissive of substantial updating from the data while providing regularisation in the presence of high input collinearity (condition number 273). Cost shares are computed by allocating actual staff costs proportionally across FTE categories, which assumes a uniform average cost per FTE across staff groups; prior standard deviations of 0.15 allow substantial updating from the data if this assumption is violated. The intercept prior is set to $\alpha \sim \mathcal{N}(\overline{\ln\text{CWA}},\, 2^2)$, centred at the sample mean of log output. Noise and baseline inefficiency variance are given weakly informative priors: $\log\sigma_v \sim \mathcal{N}(-2.5,\, 1^2)$ and $\log\sigma_{u,0} \sim \mathcal{N}(-3,\, 1^2)$. Full prior specifications for the CES and translog robustness models are provided in the Supplementary Material.

\subsection{Inefficiency Equation}

Adapting \citet{bc1995}, technical inefficiency is modelled as:

\begin{equation}
u_i \sim \mathcal{N}^+\!\left(0,\, \sigma_{u,i}^2\right), \quad
\ln \sigma_{u,i} = \ln \sigma_{u,0} + \gamma d_i +
\gamma_T T_i + \gamma_D D_i + \gamma_M m_i
\label{eq:ineff}
\end{equation}

\noindent where $d_i$ is the standardised digital maturity composite score, $T_i$ is a teaching trust indicator, $D_i$ is a financial deficit indicator, and $m_i$ is the standardised IMD score. A negative $\gamma$ implies that higher digital maturity is associated with lower inefficiency variance, that is, with greater technical efficiency. The prior is specified as $\gamma \sim \mathcal{N}(0,\, 0.5^2)$, reflecting prior ignorance about the direction and magnitude of the effect. This prior places 95\% of prior mass between $-1.0$ and $+1.0$: a value of $-1.0$ would imply that a trust one standard deviation above average in digital maturity has an inefficiency variance approximately 37\% of the baseline, which is regarded as an upper bound on plausible effect sizes. Controls are given the same weakly informative prior: $\gamma_T, \gamma_D, \gamma_M \sim \mathcal{N}(0,\, 0.5^2)$.

Under the half-normal parameterisation, the expected technical efficiency of trust $i$ is $E[\exp(-u_i)] = f(\sigma_{u,i})$, a decreasing function of $\sigma_{u,i}$. A negative $\gamma$ therefore implies that higher digital maturity is associated with lower $\sigma_{u,i}$, which in turn implies higher expected technical efficiency. Mechanistically, this is consistent with digitally mature trusts exhibiting more standardised clinical processes, reduced process variation, and fewer extreme inefficiency realisations rather than a uniform shift of all trusts toward the frontier. The variance parameterisation follows \citet{bc1995}.

\subsection{Model Variants}

Three model variants are estimated. Model A includes the composite digital maturity score in the inefficiency equation. Model B augments Model A with a dispersion term (the within-trust standard deviation across the seven pillar scores, standardised), testing whether balanced digital development matters independently of the composite level. Model C replaces the composite score with a hierarchical model over the seven DMA pillars. Pillar coefficients are drawn from a common distribution with mean $\gamma_\mu$ and standard deviation $\gamma_\sigma$, estimated from the data, so that:

\begin{equation}
\gamma_{\text{pillar},j} = \gamma_\mu + \gamma_\sigma \cdot z_j, \quad
z_j \sim \mathcal{N}(0,\, 1)
\end{equation}

\noindent This non-centred parameterisation enables partial pooling of pillar-specific estimates toward a common mean, providing regularisation when
individual pillar effects are imprecisely estimated. The prior on $\gamma_\mu$ is $\mathcal{N}(0,\, 0.3^2)$ and $\gamma_\sigma$ is given a
$\text{HalfNormal}(0.3)$ prior with a floor of 0.05 to prevent posterior degeneracy near zero.

\subsection{Estimation}

Models are estimated using Hamiltonian Monte Carlo via PyMC 6.0.1 \citep{abril2023pymc}. Four chains of 5,000 draws each are retained after 8,000 tuning iterations with target acceptance rate 0.97 (0.99 for Model C). Convergence is assessed via $\hat{R}$ statistics and effective sample size. All $\hat{R} \leq 1.010$ for parameters of interest; effective sample size for $\gamma$ exceeds 650 in all primary specifications. Log-likelihood is computed for leave-one-out cross-validation model comparison.

\subsection{Robustness}

Functional form robustness is assessed via constant elasticity of substitution (CES) and translog specifications; details are provided in the Supplementary Material. Prior sensitivity is assessed via four specifications for $\gamma$: baseline $\mathcal{N}(0, 0.5^2)$, diffuse $\mathcal{N}(0, 1.0^2)$, sceptical $\mathcal{N}(0, 0.2^2)$, and enthusiastic $\mathcal{N}(-0.1, 0.5^2)$. Capital confounding is assessed by computing Pearson correlations between physical capital (and capital intensity) and digital maturity scores across all seven DMA pillars.

\section{Results}

\subsection{Descriptive Statistics}

Descriptive statistics for the 111-trust analytical sample are presented in Table~\ref{tab:desc}. Mean CWA is \pounds 685 million; mean clinical FTE is 4,637; mean administrative FTE is 2,350; mean NLE is \pounds 353 million; and mean physical capital is \pounds 442 million. Digital maturity composite scores range from 1.67 to 3.44 with a sample mean of 2.5. Physical capital is found to be weakly and non-significantly correlated with the digital maturity composite ($r = 0.161$, $p = 0.090$) and orthogonal to six of the seven DMA pillars; one pillar (Empower People) reaches nominal significance ($r = 0.289$, $p = 0.002$) but does not survive Bonferroni correction for multiple comparisons ($p < 0.006$ threshold for 16 simultaneous tests), and one significant result in 16 tests is consistent with the expected false discovery rate under the null hypothesis. All correlations between capital intensity and DMA scores are small and non-significant. Full results including correlation heatmaps are presented in the Supplementary Material (Table~S\ref{tab:capital_corr} and Figure~S\ref{fig:capital_corr}). Capital endowment is therefore not regarded as a material confound for the efficiency premium, though the possibility of a weak association between physical capital stock and the Empower People pillar is noted.

\subsection{Production Function}

Production function estimates are stable across all three model variants (Table~\ref{tab:results}). In the primary specification (Model A), estimated output elasticities are: clinical FTE 0.423 (95\% CrI 0.25--0.59), administrative FTE 0.204 (0.09--0.32), NLE 0.264 (0.14--0.39), and capital 0.055 ($-$0.03--0.14). Returns to scale are estimated at 0.946 (95\% CrI 0.896--0.997), indicating mildly decreasing returns consistent with coordination costs at larger trust scale. The clinical FTE elasticity is approximately twice that of administrative FTE, consistent with the relative cost shares and with clinical staff being the primary productive input in acute hospital care. The capital coefficient spans zero, reflecting limited independent variation after conditioning on labour and NLE inputs (pairwise log-input correlations 0.82--0.97); its inclusion does not materially affect the digital maturity coefficient.

\subsection{Digital Maturity and Efficiency}

The composite digital maturity score is negatively associated with technical inefficiency in all three model variants. In Model A, $\hat{\gamma} = -0.612$ (95\% CrI $[-1.289, +0.005]$, $P(\gamma < 0) = 0.974$). The credible interval nearly excludes zero, with strong posterior probability concentrated below zero. Model B yields $\hat{\gamma} = -0.660$ ($P(\gamma < 0) = 0.971$); the dispersion term spans zero ($\hat{\gamma}_{\text{disp}} = -0.025$, 95\% CrI $[-0.570, +0.490]$), indicating that the balance of digital development across pillars is not found to matter independently of the composite level. LOO-CV weights are indistinguishable across Models A, B, and C (maximum ELPD difference 1.0 standard error), and Model A is preferred on parsimony.

\begin{figure}[htbp]
\centering
\includegraphics[width=\textwidth]{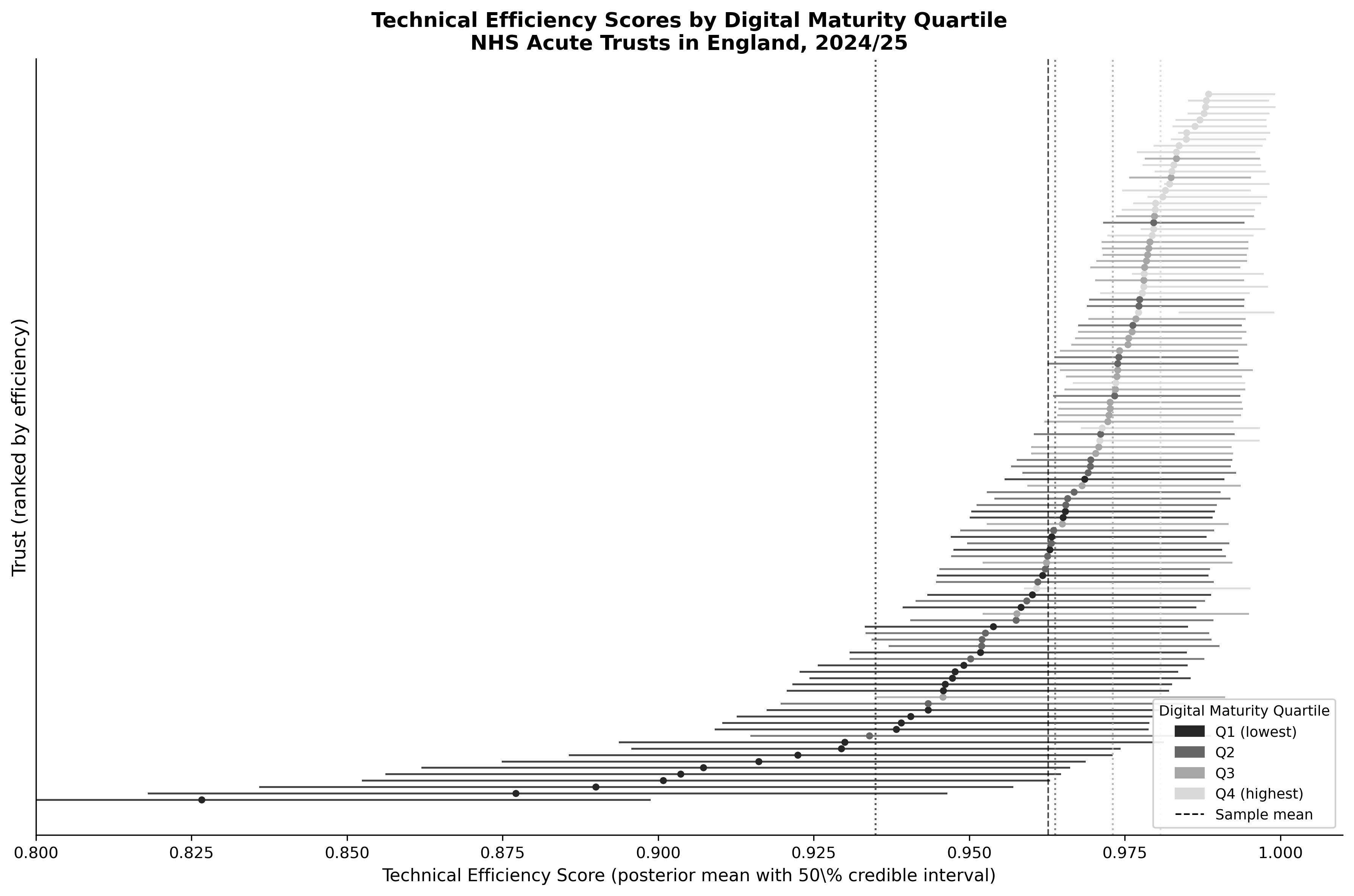}
\caption{Technical efficiency scores for 111 NHS acute non-specialist
trusts in England, 2024/25, ranked by posterior mean efficiency (Model A).
Points represent posterior mean efficiency scores with 50\% credible
intervals; shading indicates digital maturity quartile (Q1 lowest, darkest;
Q4 highest, lightest). Vertical dotted lines show quartile mean efficiency;
the dashed line shows the sample mean (96.1\%). A full ranking of all trusts
is provided in the Supplementary Material.}
\label{fig:caterpillar}
\end{figure}

Trust-level efficiency scores range from 82.0\% to 98.8\% with a sample mean of 96.1\% (standard deviation 2.6 percentage points). The Pearson correlation between the digital maturity composite and the posterior mean efficiency score is 0.711 (Figure~S\ref{fig:scatter_dma} in the Supplementary Material). Mean efficiency by DMA quartile is: Q1 93.2\%, Q2 96.2\%, Q3 97.2\%, Q4 98.0\% (Figure~\ref{fig:caterpillar}). The Q1--Q4 gap of 4.8 percentage points implies approximately \pounds 20 million of additional cost-weighted activity at mean output levels per trust, or \pounds 1.1 billion in aggregate across the lowest-quartile trusts. Catchment deprivation is not found to have an independent efficiency effect in any specification ($\hat{\gamma}_{\text{IMD}}$ spans zero in all models), indicating that the digital maturity efficiency premium is not a proxy for serving a less deprived population.

\subsection{Pillar Analysis}

In Model C, the mean pillar effect is estimated at $\hat{\gamma}_\mu = -0.211$ (95\% CrI $[-0.450, +0.019]$), confirming a consistent negative association across pillars. Pillar-specific estimates, ordered by magnitude, are presented in Table~S1 of the Supplementary Material. Healthy Populations is found to have the largest association ($\hat{\gamma} = -0.296$, $P(\gamma < 0) = 0.961$), followed by Safe Practice ($-0.266$) and Improve Care ($-0.258$, $P(\gamma < 0) = 0.931$). No individual pillar excludes zero after accounting for hierarchical shrinkage, consistent with limited power to distinguish pillar effects at $n = 111$. The consistent ordering is suggestive that population health management and care pathway optimisation capabilities have the strongest associations with efficiency, though this interpretation is exploratory given the available sample size.

\subsection{Robustness}

Robustness check results are summarised in Table~\ref{tab:robust}. Under a CES functional form, $\hat{\gamma} = -0.628$ (95\% CrI $[-1.289, -0.041]$, $P(\gamma < 0) = 0.981$), consistent with the primary specification. The substitution parameter $\rho$ spans zero (posterior mean 0.544, 95\% CrI $[-1.081, +2.104]$), confirming that the Cobb-Douglas unit elasticity restriction is not rejected. LOO-CV weights are similar across Cobb-Douglas and CES (0.53 vs 0.47), providing no evidence against the primary specification. Under a translog specification, $\hat{\gamma} = -0.630$ (95\% CrI $[-1.300, -0.059]$); LOO-CV weights again favour Cobb-Douglas
marginally (0.56 vs 0.44). Full details of the CES and translog specifications and results are provided in the Supplementary Material.

Prior sensitivity results are presented in Table~\ref{tab:robust}. The baseline, diffuse, and enthusiastic priors all yield credible intervals that exclude or nearly exclude zero. Under the sceptical prior $\mathcal{N}(0, 0.2^2)$, which constrains 95\% of prior mass to within $\pm 0.39$, the posterior mean is $-0.190$ (95\% CrI $[-0.530, +0.170]$, $P(\gamma < 0) = 0.856$), spanning zero. This prior imposes strong a priori scepticism that may be regarded as conservative for a novel empirical question where no prior evidence exists to justify such tight shrinkage.

\section{Discussion}

Evidence from this analysis indicates that digital maturity is associated with technical efficiency in NHS acute hospitals. The estimated association is negative in sign across all model specifications and functional forms, and is robust to alternative prior assumptions except under the most restrictive specification. The magnitude of the association implies that trusts in the highest digital maturity quartile operate closer to their estimated production frontier than those in the lowest quartile. These results are consistent with the hypothesis that organisational digital capabilities are correlated with
observed variation in technical efficiency.

The pillar-level analysis, while underpowered to distinguish individual effects, suggests heterogeneity in the association across domains of digital maturity. Capabilities related to population health management and care pathway optimisation are estimated to have larger negative coefficients than those related to administrative or governance functions, though credible intervals for all individual pillars include zero after hierarchical shrinkage. This pattern is consistent with mechanisms operating through clinical pathway coordination and demand management rather than through administrative substitution, but the
pillar-level results should be interpreted as exploratory given the available sample size.

A central concern in cross-sectional efficiency studies is that the variable of interest may capture unobserved organisational characteristics rather than the effect of a specific mechanism. Several aspects of the analysis reduce the plausibility of this interpretation. Physical capital, a potential proxy for organisational resources or investment environment, is not found to be systematically related to digital maturity across the DMA domains. Teaching status, financial deficit, and catchment deprivation are included directly in the inefficiency equation and are not estimated to have independent effects, suggesting that the digital maturity coefficient does not reflect these observable dimensions of organisational context. In addition, the association is stable across alternative prior specifications, including those that impose substantial shrinkage toward zero.

These considerations are not sufficient to establish causal interpretation in the absence of exogenous variation in digital maturity. Unobserved factors such as managerial capability, governance structures, or organisational culture may remain correlated with both digital maturity and efficiency. The results should
therefore be interpreted as a conditional association that is consistent with, but does not identify, a causal effect of digital capability on technical efficiency.

Several further limitations are acknowledged. The NHS DMA is a fairly new and therefore immature self-reported survey instrument. Comparison with externally validated assessments indicates that some trusts with long-standing clinical system investment may be underscored by the DMA, which would attenuate estimated effects toward zero, implying that reported estimates are likely conservative. The sceptical prior sensitivity result ($P(\gamma < 0) = 0.856$) reflects the modest sample size of 111 trusts and should be interpreted accordingly; the association is present in the data but is not sufficiently strong to dominate aggressive prior shrinkage.  inally, CWA captures acute inpatient and outpatient activity but excludes community services, research, and education, outputs that may be relevant for some trusts and that could introduce residual measurement error in the production function.

\section{Conclusion}

Digital maturity is found to be associated with technical efficiency in NHS acute hospitals. Trusts in the highest DMA quartile are estimated to operate approximately 5 percentage points closer to their production frontier than those in the lowest quartile. The association survives adjustment for input costs, physical capital, catchment deprivation, teaching status, and financial position, and is robust to functional form. The most conservative prior specification yields weaker evidence, reflecting a modest sample and the inherent difficulty of identifying organisational effects in cross-sectional data. Population health management and care pathway optimisation capabilities are found to have the strongest pillar-level associations with efficiency, though these results are exploratory. The results are consistent with digital capability being a contributor to technical efficiency in NHS acute hospitals. Causal identification would require exogenous variation in digital investment, for example from a natural experiment or staggered rollout, which is not available in this cross-sectional setting. Pending such evidence, the results nonetheless provide a quantitative benchmark for the scale of efficiency differences associated with digital maturity, and suggest that population health management and care pathway optimisation capabilities warrant particular attention in future investment prioritisation.

\newpage

\begin{table}[htbp]
\centering
\caption{Descriptive Statistics: 111 NHS Acute Non-Specialist Trusts, 2024/25}
\label{tab:desc}
\begin{threeparttable}
\begin{tabular}{lrrrrr}
\toprule
Variable & Mean & SD & Min & Median & Max \\
\midrule
\multicolumn{6}{l}{\textit{Output}} \\
CWA (\pounds m) & 685 & 391 & 156 & 573 & 2,264 \\
\midrule
\multicolumn{6}{l}{\textit{Inputs}} \\
Clinical FTE & 4,637 & 2,701 & 1,352 & 3,844 & 16,512 \\
Administrative FTE & 2,350 & 1,233 & 742 & 2,042 & 6,804 \\
Non-labour expenditure (\pounds m) & 353 & 248 & 77 & 275 & 1,351 \\
Physical capital (\pounds m) & 442 & 300 & 83 & 349 & 1,788 \\
\midrule
\multicolumn{6}{l}{\textit{Digital maturity}} \\
Composite score (1--5) & 2.50 & 0.38 & 1.67 & 2.44 & 3.44 \\
\midrule
\multicolumn{6}{l}{\textit{Controls}} \\
Teaching trust (n) & \multicolumn{5}{l}{49 of 111 (44.1\%)} \\
Financial deficit (n) & \multicolumn{5}{l}{89 of 111 (80.2\%} \\
\midrule
\multicolumn{6}{l}{\textit{Efficiency (Model A)}} \\
Technical efficiency & 0.961 & 0.026 & 0.820 & 0.969 & 0.988 \\
\bottomrule
\end{tabular}
\begin{tablenotes}
\small
\item CWA: MFF-adjusted cost-weighted activity from NCC 2024/25.
Clinical FTE: medical staff plus professionally qualified clinical
non-medical staff (ESR September 2024). Administrative FTE: support
to clinical staff plus NHS infrastructure support (ESR September 2024).
NLE: TAC total operating expenditure minus TAC actual staff costs.
Physical capital: PPE plus right-of-use assets (TAC 2024/25).
Digital maturity composite: mean of seven DMA pillar scores.
\end{tablenotes}
\end{threeparttable}
\end{table}

\newpage

\begin{table}[htbp]
\centering
\caption{Production Function and Inefficiency Equation Estimates}
\label{tab:results}
\begin{threeparttable}
\small
\begin{tabular}{lcccccc}
\toprule
& \multicolumn{2}{c}{Model A} & \multicolumn{2}{c}{Model B} &
\multicolumn{2}{c}{Model C} \\
\cmidrule(lr){2-3}\cmidrule(lr){4-5}\cmidrule(lr){6-7}
Parameter & Mean & 95\% CrI & Mean & 95\% CrI & Mean & 95\% CrI \\
\midrule
\multicolumn{7}{l}{\textit{Production function}} \\
$\beta_{\text{clinical}}$ & 0.423 & [0.25, 0.59] &
                             0.425 & [0.25, 0.60] &
                             0.417 & [0.25, 0.58] \\
$\beta_{\text{admin}}$    & 0.204 & [0.09, 0.32] &
                             0.203 & [0.09, 0.32] &
                             0.207 & [0.09, 0.32] \\
$\beta_{\text{NLE}}$      & 0.264 & [0.14, 0.39] &
                             0.262 & [0.14, 0.39] &
                             0.259 & [0.14, 0.38] \\
$\beta_{\text{capital}}$  & 0.055 & [$-$0.03, 0.14] &
                             0.056 & [$-$0.03, 0.14] &
                             0.058 & [$-$0.03, 0.14] \\
Returns to scale          & 0.946 & [0.896, 0.997] & & & & \\
\midrule
\multicolumn{7}{l}{\textit{Inefficiency equation}} \\
$\gamma_{\text{composite}}$ & $-$0.612 & [$-$1.289, $+$0.005] &
                               $-$0.660 & [$-$1.300, $+$0.034] &
                               -- & -- \\
$\gamma_{\text{dispersion}}$ & -- & -- &
                               $-$0.025 & [$-$0.570, $+$0.490] &
                               -- & -- \\
$\gamma_\mu$ (pillar mean) & -- & -- & -- & -- &
                               $-$0.211 & [$-$0.450, $+$0.019] \\
$\gamma_{\text{teaching}}$ & $-$0.090 & [$-$0.970, $+$0.680] &
                              $-$0.150 & [$-$1.000, $+$0.650] &
                              $-$0.210 & [$-$1.100, $+$0.600] \\
$\gamma_{\text{deficit}}$  & $-$0.150 & [$-$0.910, $+$0.640] &
                              $-$0.150 & [$-$0.950, $+$0.660] &
                              $-$0.230 & [$-$1.100, $+$0.620] \\
$\gamma_{\text{IMD}}$      & $-$0.089 & [$-$0.610, $+$0.380] &
                              $-$0.090 & [$-$0.630, $+$0.410] &
                              $-$0.050 & [$-$0.610, $+$0.460] \\
\midrule
\multicolumn{7}{l}{\textit{Convergence}} \\
Max $\hat{R}$ &
  \multicolumn{2}{c}{1.006} &
  \multicolumn{2}{c}{1.010} &
  \multicolumn{2}{c}{1.006} \\
ESS ($\gamma$) &
  \multicolumn{2}{c}{822} &
  \multicolumn{2}{c}{659} &
  \multicolumn{2}{c}{1,092} \\
$P(\gamma < 0)$ &
  \multicolumn{2}{c}{0.974} &
  \multicolumn{2}{c}{0.971} &
  \multicolumn{2}{c}{--} \\
\bottomrule
\end{tabular}
\begin{tablenotes}
\small
\item Bayesian estimation via HMC (PyMC 6.0.1). Four chains, 5,000
draws, 8,000 tuning iterations, target accept 0.97 (0.99 for Model C).
CrI: 95\% credible interval. Model A: composite digital maturity.
Model B: composite plus dispersion. Model C: hierarchical pillar model
(seven pillars, partial pooling). $P(\gamma < 0)$: posterior probability
of negative composite effect. Negative $\gamma$ indicates higher digital
maturity associated with lower inefficiency. ESS reported for primary
digital maturity parameter.
\end{tablenotes}
\end{threeparttable}
\end{table}

\newpage

\begin{table}[htbp]
\centering
\caption{Robustness Checks}
\label{tab:robust}
\begin{threeparttable}
\begin{tabular}{llrrrr}
\toprule
Check & Specification & $\hat{\gamma}$ & 95\% CrI & $P(\gamma<0)$ &
LOO weight \\
\midrule
\multicolumn{6}{l}{\textit{Functional form}} \\
& Cobb-Douglas (primary) & $-$0.612 & [$-$1.289, $+$0.005] & 0.974 & 0.53 \\
& CES                    & $-$0.628 & [$-$1.289, $-$0.041] & 0.981 & 0.47 \\
& Translog               & $-$0.630 & [$-$1.300, $-$0.059] & --    & 0.44 \\
\midrule
\multicolumn{6}{l}{\textit{Prior sensitivity ($\gamma_{\text{composite}}$)}} \\
& Baseline $\mathcal{N}(0, 0.5^2)$        &
  $-$0.612 & [$-$1.289, $+$0.005] & 0.974 & -- \\
& Diffuse $\mathcal{N}(0, 1.0^2)$         &
  $-$1.040 & [$-$1.900, $-$0.210] & 0.989 & -- \\
& Enthusiastic $\mathcal{N}(-0.1, 0.5^2)$ &
  $-$0.690 & [$-$1.300, $-$0.053] & 0.985 & -- \\
& Sceptical $\mathcal{N}(0, 0.2^2)$       &
  $-$0.190 & [$-$0.530, $+$0.170] & 0.856 & -- \\
\midrule
\multicolumn{6}{l}{\textit{Input specification}} \\
& Looser $\beta$ priors ($\sigma = 0.5$) &
  $-$0.649 & [$-$1.307, $-$0.001] & 0.975 & -- \\
& Three-input (capital excluded)          &
  $-$0.614 & [$-$1.285, $+$0.003] & 0.974 & -- \\
\bottomrule
\end{tabular}
\begin{tablenotes}
\small
\item CES: constant elasticity of substitution; substitution parameter
$\rho$ posterior mean 0.544 (95\% CrI $[-1.081, +2.104]$); Cobb-Douglas
restriction not rejected. LOO weights from PSIS-LOO-CV. Translog LOO weight shown relative to
Cobb-Douglas (0.56); all functional form comparisons are
indistinguishable within 1 standard error of ELPD. Translog:
$\hat{R} = 1.01$; interpreted with caution. Prior sensitivity reruns
Model A with alternative priors on $\gamma_{\text{composite}}$ using
2,000 draws. Sceptical prior constrains 95\% of prior mass to within
$\pm 0.39$; credible interval spans zero under this specification.
Input specification checks use the primary prior and sampling settings;
efficiency rank correlations with the primary specification are 0.996
(looser $\beta$ priors) and 0.996 (capital excluded), confirming that
neither the production function prior nor the inclusion of weakly
identified capital materially affects the efficiency ranking or the
digital maturity coefficient. Supplementary Material provides full
details of CES and translog specifications.
\end{tablenotes}
\end{threeparttable}
\end{table}

\newpage

\newpage

\bibliographystyle{apalike}
\bibliography{references}

@article{bc1995,
  author  = {Battese, George E. and Coelli, Timothy J.},
  title   = {A model for technical inefficiency effects in a stochastic
             frontier production function for panel data},
  journal = {Empirical Economics},
  year    = {1995},
  volume  = {20},
  number  = {2},
  pages   = {325--332},
  doi     = {10.1007/BF01205442}
}

@article{castelli2011,
  author  = {Castelli, Adriana and Laudicella, Mauro and Street, Andrew
             and Ward, Padraic},
  title   = {Getting out what we put in: productivity of the {English}
             {National Health Service}},
  journal = {Health Economics, Policy and Law},
  year    = {2011},
  volume  = {6},
  number  = {3},
  pages   = {313--335},
  doi     = {10.1017/S1744133110000211}
}

@article{bardhan2013electronic,
  author  = {Bardhan, Indranil R. and Thouin, Mark F.},
  title   = {Health information technology and its impact on the quality
             and cost of healthcare delivery},
  journal = {Decision Support Systems},
  year    = {2013},
  volume  = {55},
  number  = {2},
  pages   = {438--449},
  doi     = {10.1016/j.dss.2012.10.003}
}

@article{abril2023pymc,
  author  = {Abril-Pla, Oriol and Andreani, Virgile and Carroll, Colin
             and Dong, Larry and Fonnesbeck, Christopher J. and
             Kochurov, Maxim and Kumar, Ravin and Lao, Junpeng and
             Luhmann, Christian C. and Martin, Osvaldo A. and
             Osthege, Michael and Vieira, Ricardo and Wiecki, Thomas
             and Zinkov, Robert},
  title   = {{PyMC}: a modern, and comprehensive probabilistic
             programming framework in {Python}},
  journal = {PeerJ Computer Science},
  year    = {2023},
  volume  = {9},
  pages   = {e1516},
  doi     = {10.7717/peerj-cs.1516}
}

@misc{nhse2021digital,
  author       = {{NHS England}},
  title        = {What Good Looks Like: Digital Transformation Framework},
  year         = {2021},
  howpublished = {\url{https://www.england.nhs.uk/digitaltechnology/what-good-looks-like}}
}

@misc{nhsi2019tech,
  author       = {{NHS Improvement}},
  title        = {Releasing Time to Care: The Productive Ward and
                  Digital Technology},
  year         = {2019},
  howpublished = {\url{https://www.england.nhs.uk}}
}

@misc{nhse2025ncc,
  author       = {{NHS England}},
  title        = {National Cost Collection 2024/25},
  year         = {2025},
  howpublished = {\url{https://www.england.nhs.uk/national-cost-collection}}
}

@misc{nhse2025esr,
  author       = {{NHS England}},
  title        = {{NHS} Workforce Statistics, {September} 2024},
  year         = {2025},
  howpublished = {\url{https://digital.nhs.uk/data-and-information/publications/statistical/nhs-workforce-statistics}}
}

@misc{nhse2025tac,
  author       = {{NHS England}},
  title        = {{NHS} Trust Annual Accounts 2024/25},
  year         = {2025},
  howpublished = {\url{https://www.england.nhs.uk/financial-accounting-and-reporting/nhs-provider-accounts}}
}

@misc{nhse2025dma,
  author       = {{NHS England}},
  title        = {Digital Maturity Assessment 2025},
  year         = {2025},
  howpublished = {\url{https://www.england.nhs.uk/digitaltechnology/connecteddigitalsystems/digital-maturity}}
}

@misc{nhse2026imd,
  author       = {{NHS England}},
  title        = {{NHS} Acute Hospital Trust Catchment Populations,
                  {April} 2026},
  year         = {2026},
  howpublished = {\url{https://www.england.nhs.uk/publication/nhs-acute-hospital-trust-catchment-populations}}
}

@misc{nhse2025oversight,
  author       = {{NHS England}},
  title        = {{NHS} Oversight Framework: Acute Trust League Table
                  2024/25},
  year         = {2025},
  howpublished = {\url{https://www.england.nhs.uk/publication/nhs-system-oversight-framework}}
}

\newpage

\section*{Data Availability}

All data used in this study are publicly available. Cost-weighted activity
and MFF-adjusted expenditure: NHS National Cost Collection 2024/25
(\url{https://www.england.nhs.uk/national-cost-collection}).
Workforce data: NHS Workforce Statistics, September 2024
(\url{https://digital.nhs.uk/data-and-information/publications/statistical/nhs-workforce-statistics}).
Financial data: NHS Trust Annual Accounts 2024/25
(\url{https://www.england.nhs.uk/financial-accounting-and-reporting/nhs-provider-accounts}).
Digital Maturity Assessment: NHS England
(\url{https://www.england.nhs.uk/digitaltechnology/connecteddigitalsystems/digital-maturity}).
Catchment IMD scores: NHS Acute Hospital Trust Catchment Populations,
April 2026
(\url{https://www.england.nhs.uk/publication/nhs-acute-hospital-trust-catchment-populations}).
NHS Oversight Framework: NHS England
(\url{https://www.england.nhs.uk/publication/nhs-system-oversight-framework}).

\section*{Conflict of Interest}

No conflicts of interest are declared.

\section*{Funding}

This research received no specific funding.

\section*{AI Disclosure}

Large language model assistance (Claude AI Sonnet 4.6) was used to assist
the author with manuscript typesetting (LaTeX generation) and to perform
code double-checking. AI was not used in the design of this work and did not
influence the results or interpretation.

\newpage
\section*{Supplementary Material}
\setcounter{section}{0}
\renewcommand{\thesection}{S\arabic{section}}
\renewcommand{\thetable}{S\arabic{table}}
\renewcommand{\thefigure}{S\arabic{figure}}
\setcounter{table}{0}
\setcounter{figure}{0}







\maketitle
\thispagestyle{empty}

\setcounter{page}{1}

\section{Prior Specification}

Full prior distributions for all model parameters are listed in
Table~\ref{tab:priors}. The rationale for production function priors is
described in the main text. All models use the same priors unless otherwise
noted.

\begin{sidewaystable}[htbp]
\centering
\caption{Prior Distributions}
\label{tab:priors}
\begin{threeparttable}
\begin{tabular}{llp{10cm}}
\toprule
Parameter & Prior & Rationale \\
\midrule
\multicolumn{3}{l}{\textit{Production function (all models)}} \\
$\alpha$ & $\mathcal{N}(\overline{\ln\text{CWA}},\, 2^2)$ &
  Weakly informative, centred at mean log output \\
$\beta_{\text{clinical}}$ & $\mathcal{N}(0.378,\, 0.15^2)$ &
  TAC-derived cost share \\
$\beta_{\text{admin}}$ & $\mathcal{N}(0.197,\, 0.15^2)$ &
  TAC-derived cost share \\
$\beta_{\text{NLE}}$ & $\mathcal{N}(0.376,\, 0.15^2)$ &
  TAC-derived cost share \\
$\beta_{\text{capital}}$ & $\mathcal{N}(0.049,\, 0.10^2)$ &
  TAC-derived cost share (tighter: small share) \\
$\log\sigma_v$ & $\mathcal{N}(-2.5,\, 1^2)$ & Weakly informative \\
$\log\sigma_{u,0}$ & $\mathcal{N}(-3,\, 1^2)$ & Weakly informative \\
\midrule
\multicolumn{3}{l}{\textit{Inefficiency equation (Models A and B)}} \\
$\gamma_{\text{composite}}$ & $\mathcal{N}(0,\, 0.5^2)$ &
  Prior ignorance on direction and magnitude \\
$\gamma_{\text{dispersion}}$ (Model B) & $\mathcal{N}(0,\, 0.5^2)$ &
  Weakly informative \\
$\gamma_{\text{teaching}},\, \gamma_{\text{deficit}},\,
  \gamma_{\text{IMD}}$ & $\mathcal{N}(0,\, 0.5^2)$ & Weakly informative \\
\midrule
\multicolumn{3}{l}{\textit{Hierarchical pillar model (Model C)}} \\
$\gamma_\mu$ & $\mathcal{N}(0,\, 0.3^2)$ &
  Slightly tighter than composite prior \\
$\gamma_{\sigma,\text{raw}}$ & $\text{HalfNormal}(0.3)$ &
  Dispersion of pillar effects; floored at 0.05 \\
$\gamma_\sigma$ & $\gamma_{\sigma,\text{raw}} + 0.05$ &
  Floor prevents posterior degeneracy near zero \\
$z_j$ & $\mathcal{N}(0,\, 1)$ &
  Non-centred parameterisation for numerical stability \\
$\gamma_{\text{teaching}},\, \gamma_{\text{deficit}},\,
  \gamma_{\text{IMD}}$ & $\mathcal{N}(0,\, 0.5^2)$ & Weakly informative \\
\midrule
\multicolumn{3}{l}{\textit{CES model}} \\
$\delta_1, \delta_2, \delta_3, \delta_4$ & $\text{Dirichlet}(10 \times
  [0.378, 0.197, 0.376, 0.049])$ & TAC cost shares as concentration \\
$\rho_{\text{raw}}$ & $\mathcal{N}(0,\, 1^2)$ &
  Weakly informative on substitution parameter \\
$\nu$ & $\mathcal{N}(1,\, 0.1^2)$ &
  Returns to scale prior centred on unity \\
\midrule
\multicolumn{3}{l}{\textit{Translog model (additional parameters)}} \\
$g_{cc}, g_{aa}, g_{nn}, g_{kk}$ & $\mathcal{N}(0,\, 0.1^2)$ &
  Second-order own terms \\
$g_{ca}, g_{cn}, g_{ck}, g_{an}, g_{ak}, g_{nk}$ &
  $\mathcal{N}(0,\, 0.1^2)$ & Cross terms \\
\bottomrule
\end{tabular}
\begin{tablenotes}
\small
\item Cost shares are computed from NHS Trust Annual Accounts 2024/25
as the ratio of each input's cost to total operating expenditure,
averaged across the 111-trust analytical sample.
$\overline{\ln\text{CWA}}$ denotes the sample mean of log
cost-weighted activity.
\end{tablenotes}
\end{threeparttable}
\end{sidewaystable}

\section{CES Production Function}

The CES production function is specified as:

\begin{equation}
\ln(\text{CWA}_i) = \alpha + \left(-\frac{\nu}{\rho}\right)
\ln\!\left(\sum_{k=1}^{4} \delta_k \tilde{x}_{k,i}^{-\rho}\right) + v_i - u_i
\end{equation}

\noindent where $\tilde{x}_{k,i} = x_{k,i} / \bar{x}_k$ denotes each input
normalised to unit sample mean, $\delta_k$ is the input share parameter
(constrained via a Dirichlet prior to sum to unity), $\rho$ is the
substitution parameter, and $\nu$ is the returns to scale parameter. The
elasticity of substitution is $\sigma = 1/(1+\rho)$; the Cobb-Douglas case
corresponds to $\rho = 0$ ($\sigma = 1$). To avoid the discontinuity at
$\rho = 0$, a small epsilon of 0.01 is added to $|\rho_{\text{raw}}|$ after
sampling from $\rho_{\text{raw}} \sim \mathcal{N}(0,\, 1^2)$.

The inefficiency equation is identical to Model A. Estimation uses 5,000
draws after 8,000 tuning iterations per chain (four chains), target acceptance
rate 0.97.

CES results are presented in Table~\ref{tab:ces}. The digital maturity
coefficient is $\hat{\gamma} = -0.633$ (95\% CrI $[-1.322, -0.040]$,
$P(\gamma < 0) = 0.982$), essentially identical to the primary Cobb-Douglas
specification. The substitution parameter posterior mean is $\hat{\rho} = 0.546$
(95\% CrI $[-1.049, +2.094]$), spanning zero and confirming that the
Cobb-Douglas unit elasticity restriction is not rejected. Efficiency scores are
very highly correlated with those from the primary specification ($r = 0.999$).
The Cobb-Douglas model is preferred on PSIS-LOO-CV (weight 0.68 vs 0.32 for CES).

\begin{table}[htbp]
\centering
\caption{CES Model Results}
\label{tab:ces}
\begin{threeparttable}
\begin{tabular}{lrrrr}
\toprule
Parameter & Mean & SD & 95\% CrI lower & 95\% CrI upper \\
\midrule
$\delta_{\text{clinical}}$ & 0.493 & 0.099 & 0.30 & 0.68 \\
$\delta_{\text{admin}}$    & 0.194 & 0.063 & 0.07 & 0.32 \\
$\delta_{\text{NLE}}$      & 0.276 & 0.067 & 0.15 & 0.41 \\
$\delta_{\text{capital}}$  & 0.037 & 0.033 & 0.00 & 0.12 \\
$\rho$                     & 0.546 & 0.801 & $-$1.049 & $+$2.094 \\
$\nu$                      & 0.946 & 0.025 & 0.90 & 1.00 \\
$\gamma_{\text{composite}}$& $-$0.633 & 0.340 & $-$1.322 & $-$0.040 \\
$\gamma_{\text{IMD}}$      & $-$0.091 & 0.235 & $-$0.589 & $+$0.376 \\
\midrule
$P(\gamma_{\text{composite}} < 0)$ & \multicolumn{4}{l}{0.982} \\
Efficiency correlation with CD     & \multicolumn{4}{l}{0.999} \\
LOO weight (vs Cobb-Douglas 0.68)  & \multicolumn{4}{l}{0.32} \\
\bottomrule
\end{tabular}
\begin{tablenotes}
\small
\item Teaching trust, financial deficit, and IMD coefficients are
omitted for brevity; all span zero. CD: Cobb-Douglas primary model.
\end{tablenotes}
\end{threeparttable}
\end{table}

\section{Translog Production Function}

The translog production function is specified as:

\begin{align}
\ln(\text{CWA}_i) = \; &\alpha
  + \beta_c \widetilde{\ln x}_{c,i}
  + \beta_a \widetilde{\ln x}_{a,i}
  + \beta_n \widetilde{\ln x}_{n,i}
  + \beta_k \widetilde{\ln x}_{k,i} \nonumber \\
&+ \tfrac{1}{2} g_{cc} \widetilde{\ln x}_{c,i}^2
  + \tfrac{1}{2} g_{aa} \widetilde{\ln x}_{a,i}^2
  + \tfrac{1}{2} g_{nn} \widetilde{\ln x}_{n,i}^2
  + \tfrac{1}{2} g_{kk} \widetilde{\ln x}_{k,i}^2 \nonumber \\
&+ g_{ca} \widetilde{\ln x}_{c,i} \widetilde{\ln x}_{a,i}
  + g_{cn} \widetilde{\ln x}_{c,i} \widetilde{\ln x}_{n,i}
  + g_{ck} \widetilde{\ln x}_{c,i} \widetilde{\ln x}_{k,i} \nonumber \\
&+ g_{an} \widetilde{\ln x}_{a,i} \widetilde{\ln x}_{n,i}
  + g_{ak} \widetilde{\ln x}_{a,i} \widetilde{\ln x}_{k,i}
  + g_{nk} \widetilde{\ln x}_{n,i} \widetilde{\ln x}_{k,i}
  + v_i - u_i
\end{align}

\noindent where $\widetilde{\ln x}$ denotes mean-centred log inputs. The
translog nests Cobb-Douglas when all second-order terms equal zero.
Centring is essential for numerical stability of the translog; the prior
on the intercept is set to the sample mean of log output, consistent with
the primary model. Second-order terms are given tighter priors
$\mathcal{N}(0,\, 0.1^2)$ to regularise the additional parameters. The
inefficiency equation is identical to Model A.

Estimation uses 5,000 draws after 8,000 tuning iterations per chain (four
chains), target acceptance rate 0.99. All $\hat{R} \leq 1.01$.

Translog results are presented in Table~\ref{tab:translog}. The digital
maturity coefficient is $\hat{\gamma} = -0.660$ (95\% CrI $[-1.300, -0.073]$),
excluding zero and consistent with the primary specification. First-order
coefficients are comparable to those of the primary model. The Cobb-Douglas
model is marginally preferred on PSIS-LOO-CV (weight 0.60 vs 0.40 for translog).
The translog is likely overparameterised at $n = 111$ with 15 production
function parameters; the LOO-CV preference for the more parsimonious
Cobb-Douglas specification is consistent with this interpretation.

\begin{table}[htbp]
\centering
\caption{Translog Model Results (selected parameters)}
\label{tab:translog}
\begin{threeparttable}
\begin{tabular}{lrrrr}
\toprule
Parameter & Mean & SD & 95\% CrI lower & 95\% CrI upper \\
\midrule
\multicolumn{5}{l}{\textit{First-order terms}} \\
$\beta_{\text{clinical}}$ & 0.453 & 0.088 & 0.28 & 0.63 \\
$\beta_{\text{admin}}$    & 0.178 & 0.060 & 0.06 & 0.30 \\
$\beta_{\text{NLE}}$      & 0.266 & 0.061 & 0.14 & 0.39 \\
$\beta_{\text{capital}}$  & 0.054 & 0.041 & $-$0.03 & 0.13 \\
\midrule
\multicolumn{5}{l}{\textit{Inefficiency equation}} \\
$\gamma_{\text{composite}}$ & $-$0.660 & 0.318 & $-$1.300 & $-$0.073 \\
\midrule
$\hat{R}$ (max) &
  \multicolumn{4}{l}{1.01} \\
LOO weight (vs Cobb-Douglas 0.60) &
  \multicolumn{4}{l}{0.40} \\
\bottomrule
\end{tabular}
\begin{tablenotes}
\small
\item Second-order terms are omitted for brevity; all span zero.
Teaching trust, financial deficit, and IMD coefficients span zero
and are omitted.
\end{tablenotes}
\end{threeparttable}
\end{table}

\section{Pillar-Level Results (Model C)}

Table~\ref{tab:pillars} presents pillar-specific inefficiency coefficients
from Model C. All seven pillar coefficients are estimated to be negative,
consistent with the composite result from Model A. The mean pillar effect
$\hat{\gamma}_\mu = -0.212$ (95\% CrI $[-0.450, +0.029]$) nearly excludes
zero. No individual pillar excludes zero at 95\%, reflecting both the
modest sample size and the partial pooling induced by the hierarchical prior.
Healthy Populations and Improve Care are estimated to have the largest
posterior probabilities of a negative effect.

\begin{table}[htbp]
\centering
\caption{Pillar-Specific Inefficiency Coefficients (Model C)}
\label{tab:pillars}
\begin{threeparttable}
\begin{tabular}{lrrrrr}
\toprule
Pillar & Mean & SD & 95\% CrI lower & 95\% CrI upper & $P(\gamma<0)$ \\
\midrule
Healthy Populations  & $-$0.296 & 0.193 & $-$0.740 & $+$0.041 & 0.961 \\
Safe Practice        & $-$0.269 & 0.198 & $-$0.710 & $+$0.094 & --    \\
Improve Care         & $-$0.256 & 0.195 & $-$0.690 & $+$0.100 & 0.931 \\
Support Workforce    & $-$0.226 & 0.189 & $-$0.610 & $+$0.160 & --    \\
Empower People       & $-$0.191 & 0.186 & $-$0.550 & $+$0.190 & --    \\
Smart Foundations    & $-$0.189 & 0.200 & $-$0.580 & $+$0.230 & --    \\
Well Led             & $-$0.172 & 0.208 & $-$0.560 & $+$0.270 & --    \\
\midrule
$\gamma_\mu$ (mean)         &
  $-$0.212 & 0.120 & $-$0.450 & $+$0.029 & -- \\
$\gamma_\sigma$ (dispersion) &
  $+$0.190 & 0.116 & $+$0.055 & $+$0.480 & -- \\
\bottomrule
\end{tabular}
\begin{tablenotes}
\small
\item Negative coefficient indicates higher pillar score is associated
with lower inefficiency. $P(\gamma < 0)$ is reported only for pillars
with the two highest posterior probabilities of a negative effect.
Partial pooling toward $\gamma_\mu$ means individual pillar estimates
are shrunk toward the common mean; no individual pillar excludes zero
at the 95\% credible level.
\end{tablenotes}
\end{threeparttable}
\end{table}

\section{Trust-Level Efficiency Rankings}

Table~\ref{tab:rankings} presents posterior mean technical efficiency scores
for all 111 trusts in the analytical sample, ranked from most to least
efficient. Digital maturity composite scores are shown alongside efficiency
estimates to illustrate the association documented in the main text.

\section{Trust-Level Efficiency Rankings}

Table~\ref{tab:rankings} presents posterior mean technical efficiency scores
for all 111 trusts in the analytical sample, ranked from most to least
efficient.

\begin{center}
\small
\begin{longtable}{rlllrr}
\caption{Trust-Level Technical Efficiency Rankings (Model A)}
\label{tab:rankings} \\
\toprule
Rank & Trust & Subtype & Region & DM & Eff (\%) \\
\midrule
\endfirsthead
\multicolumn{6}{l}{\small\textit{Table~\ref{tab:rankings} continued}} \\
\toprule
Rank & Trust & Subtype & Region & DM & Eff (\%) \\
\midrule
\endhead
\midrule
\multicolumn{6}{r}{\small\textit{Continued on next page}} \\
\endfoot
\bottomrule
\multicolumn{6}{l}{\small\textit{FT: NHS Foundation Trust. DM: digital maturity composite score (scale 1--5).}} \\
\multicolumn{6}{l}{\small\textit{Eff: posterior mean technical efficiency as percentage of production frontier.}} \\
\endlastfoot
1 & South Tyneside and Sunderland NHS FT & Large & North East & 3.44 & 98.8 \\
2 & The Dudley Group NHS FT & Medium & Midlands & 3.11 & 98.8 \\
3 & Manchester University NHS FT & Teaching & North West & 3.41 & 98.8 \\
4 & Barts Health NHS Trust & Teaching & London & 3.00 & 98.8 \\
5 & North Tees and Hartlepool NHS FT & Medium & North East & 3.00 & 98.7 \\
6 & Wirral University Teaching Hospital NHS FT & Teaching & North West & 2.91 & 98.6 \\
7 & Chelsea and Westminster Hospital NHS FT & Teaching & London & 3.10 & 98.5 \\
8 & The Hillingdon Hospitals NHS FT & Small & London & 3.06 & 98.5 \\
9 & Warrington and Halton Teaching Hospitals NHS FT & Teaching & North West & 2.84 & 98.4 \\
10 & Surrey and Sussex Healthcare NHS Trust & Medium & South East & 2.76 & 98.3 \\
11 & South Tees Hospitals NHS FT & Teaching & North East & 2.67 & 98.3 \\
12 & Ashford and St Peter's Hospitals NHS FT & Medium & South East & 2.96 & 98.3 \\
13 & Mersey and West Lancashire Teaching Hospitals & Teaching & North West & 2.90 & 98.3 \\
14 & Medway NHS Foundation Trust & Medium & South East & 2.56 & 98.2 \\
15 & Imperial College Healthcare NHS Trust & Teaching & London & 3.13 & 98.2 \\
16 & East and North Hertfordshire NHS Trust & Large & East & 2.73 & 98.1 \\
17 & West Hertfordshire Teaching Hospitals NHS Trust & Teaching & East & 3.04 & 98.1 \\
18 & The Rotherham NHS FT & Small & North East & 2.87 & 98.0 \\
19 & Calderdale and Huddersfield NHS FT & Large & North East & 2.76 & 98.0 \\
20 & Doncaster and Bassetlaw Teaching Hospitals NHS FT & Teaching & North East & 2.59 & 98.0 \\
21 & Royal Free London NHS FT & Teaching & London & 2.24 & 98.0 \\
22 & Homerton Healthcare NHS FT & Teaching & London & 2.89 & 98.0 \\
23 & Frimley Health NHS FT & Large & South East & 2.76 & 97.9 \\
24 & Barnsley Hospital NHS FT & Small & North East & 2.59 & 97.9 \\
25 & University Hospitals of Derby and Burton NHS FT & Teaching & Midlands & 2.53 & 97.9 \\
26 & Worcestershire Acute Hospitals NHS Trust & Large & Midlands & 2.64 & 97.9 \\
27 & South Warwickshire University NHS FT & Teaching & Midlands & 2.60 & 97.8 \\
28 & Mid and South Essex NHS FT & Large & East & 2.44 & 97.8 \\
29 & University Hospitals Coventry and Warwickshire & Teaching & Midlands & 2.91 & 97.8 \\
30 & Lewisham and Greenwich NHS Trust & Large & London & 2.67 & 97.8 \\
31 & London North West University Healthcare NHS Trust & Teaching & London & 3.10 & 97.8 \\
32 & West Suffolk NHS FT & Small & East & 2.74 & 97.8 \\
33 & East Kent Hospitals University NHS FT & Teaching & South East & 2.43 & 97.7 \\
34 & Hull University Teaching Hospitals NHS Trust & Teaching & North East & 2.40 & 97.7 \\
35 & University College London Hospitals NHS FT & Teaching & London & 3.39 & 97.7 \\
36 & Northampton General Hospital NHS Trust & Medium & Midlands & 2.66 & 97.7 \\
37 & York and Scarborough Teaching Hospitals NHS FT & Teaching & North East & 2.41 & 97.6 \\
38 & Walsall Healthcare NHS Trust & Small & Midlands & 2.56 & 97.6 \\
39 & East Suffolk and North Essex NHS FT & Large & East & 2.64 & 97.6 \\
40 & Royal Surrey County Hospital NHS FT & Medium & South East & 2.67 & 97.5 \\
41 & Gloucestershire Hospitals NHS FT & Large & South West & 2.61 & 97.4 \\
42 & Mid Yorkshire Teaching NHS Trust & Teaching & North East & 2.31 & 97.4 \\
43 & East Lancashire Hospitals NHS Trust & Large & North West & 2.33 & 97.4 \\
44 & Bradford Teaching Hospitals NHS FT & Teaching & North East & 2.47 & 97.4 \\
45 & Royal Cornwall Hospitals NHS Trust & Large & South West & 2.67 & 97.4 \\
46 & Northumbria Healthcare NHS FT & Large & North East & 2.74 & 97.4 \\
47 & Sheffield Teaching Hospitals NHS FT & Teaching & North East & 2.50 & 97.3 \\
48 & University Hospitals Birmingham NHS FT & Teaching & Midlands & 2.33 & 97.3 \\
49 & Mid Cheshire Hospitals NHS FT & Small & North West & 2.66 & 97.3 \\
50 & The Royal Wolverhampton NHS Trust & Large & Midlands & 2.60 & 97.3 \\
51 & Sherwood Forest Hospitals NHS FT & Medium & Midlands & 2.59 & 97.2 \\
52 & Royal Devon University Healthcare NHS FT & Large & South West & 2.51 & 97.2 \\
53 & Cambridge University Hospitals NHS FT & Teaching & East & 2.94 & 97.1 \\
54 & The Newcastle Upon Tyne Hospitals NHS FT & Teaching & North East & 2.36 & 97.1 \\
55 & Countess of Chester Hospital NHS FT & Small & North West & 3.00 & 97.1 \\
56 & Wrightington, Wigan and Leigh NHS FT & Medium & North West & 2.47 & 97.1 \\
57 & George Eliot Hospital NHS Trust & Small & Midlands & 2.54 & 97.0 \\
58 & Norfolk and Norwich University Hospitals NHS FT & Teaching & East & 2.36 & 96.9 \\
59 & Bolton NHS Foundation Trust & Medium & North West & 2.39 & 96.9 \\
60 & University Hospitals of Leicester NHS Trust & Teaching & Midlands & 2.41 & 96.9 \\
61 & Barking, Havering and Redbridge University Hospitals & Teaching & London & 2.20 & 96.9 \\
62 & Leeds Teaching Hospitals NHS Trust & Teaching & North East & 2.49 & 96.8 \\
63 & Northern Lincolnshire and Goole NHS FT & Medium & North East & 2.29 & 96.7 \\
64 & University Hospital Southampton NHS FT & Teaching & South East & 2.41 & 96.6 \\
65 & Chesterfield Royal Hospital NHS FT & Small & Midlands & 2.33 & 96.5 \\
66 & Kingston and Richmond NHS FT & Medium & London & 2.10 & 96.5 \\
67 & Dorset County Hospital NHS FT & Small & South West & 2.17 & 96.5 \\
68 & Maidstone and Tunbridge Wells NHS Trust & Large & South East & 2.61 & 96.5 \\
69 & Kettering General Hospital NHS FT & Small & Midlands & 2.34 & 96.4 \\
70 & Bedfordshire Hospitals NHS FT & Medium & East & 2.00 & 96.3 \\
71 & Blackpool Teaching Hospitals NHS FT & Teaching & North West & 2.36 & 96.3 \\
72 & James Paget University Hospitals NHS FT & Teaching & East & 2.23 & 96.3 \\
73 & Sandwell and West Birmingham Hospitals NHS Trust & Large & Midlands & 2.24 & 96.3 \\
74 & Nottingham University Hospitals NHS Trust & Teaching & Midlands & 2.46 & 96.2 \\
75 & Royal Berkshire NHS FT & Large & South East & 2.24 & 96.2 \\
76 & Milton Keynes University Hospital NHS FT & Teaching & East & 2.07 & 96.2 \\
77 & Harrogate and District NHS FT & Small & North East & 2.41 & 96.1 \\
78 & Guy's and St Thomas' NHS FT & Teaching & London & 2.79 & 96.1 \\
79 & Epsom and St Helier University Hospitals NHS Trust & Teaching & London & 2.21 & 96.0 \\
80 & Portsmouth Hospitals University NHS Trust & Large & South East & 2.24 & 95.9 \\
81 & North West Anglia NHS FT & Large & East & 2.04 & 95.8 \\
82 & Liverpool University Hospitals NHS FT & Teaching & North West & 2.61 & 95.8 \\
83 & University Hospitals Dorset NHS FT & Teaching & South West & 2.36 & 95.8 \\
84 & University Hospitals Sussex NHS FT & Teaching & South East & 1.90 & 95.4 \\
85 & Oxford University Hospitals NHS FT & Teaching & South East & 2.30 & 95.3 \\
86 & University Hospitals of North Midlands NHS Trust & Teaching & Midlands & 2.37 & 95.2 \\
87 & Lancashire Teaching Hospitals NHS FT & Teaching & North West & 2.40 & 95.2 \\
88 & Stockport NHS Foundation Trust & Medium & North West & 2.23 & 95.2 \\
89 & St George's University Hospitals NHS FT & Teaching & London & 2.29 & 95.0 \\
90 & Tameside and Glossop Integrated Care NHS FT & Small & North West & 2.14 & 94.9 \\
91 & Gateshead Health NHS FT & Medium & North East & 2.06 & 94.8 \\
92 & University Hospitals of Morecambe Bay NHS FT & Teaching & North West & 2.17 & 94.7 \\
93 & East Sussex Healthcare NHS Trust & Large & South East & 2.10 & 94.6 \\
94 & Hampshire Hospitals NHS FT & Large & South East & 2.10 & 94.6 \\
95 & King's College Hospital NHS FT & Teaching & London & 2.51 & 94.6 \\
96 & Royal United Hospitals Bath NHS FT & Medium & South West & 2.36 & 94.3 \\
97 & Dartford and Gravesham NHS Trust & Small & South East & 2.01 & 94.3 \\
98 & Croydon Health Services NHS Trust & Medium & London & 2.00 & 94.1 \\
99 & Great Western Hospitals NHS FT & Medium & South West & 2.09 & 93.9 \\
100 & United Lincolnshire Hospitals NHS Trust & Large & Midlands & 2.03 & 93.8 \\
101 & University Hospitals Plymouth NHS Trust & Teaching & South West & 2.43 & 93.4 \\
102 & Northern Care Alliance NHS FT & Teaching & North West & 1.93 & 93.0 \\
103 & The Princess Alexandra Hospital NHS Trust & Small & East & 1.97 & 92.9 \\
104 & Airedale NHS Foundation Trust & Small & North East & 2.09 & 92.2 \\
105 & The Shrewsbury and Telford Hospital NHS Trust & Medium & Midlands & 1.99 & 91.6 \\
106 & North Bristol NHS Trust & Large & South West & 2.14 & 90.7 \\
107 & North Cumbria Integrated Care NHS FT & Medium & North East & 2.04 & 90.4 \\
108 & University Hospitals Bristol and Weston NHS FT & Teaching & South West & 1.91 & 90.1 \\
109 & Salisbury NHS Foundation Trust & Small & South West & 2.07 & 89.0 \\
110 & The Queen Elizabeth Hospital, King's Lynn, NHS FT & Small & East & 1.81 & 87.7 \\
111 & East Cheshire NHS Trust & Small & North West & 1.67 & 82.7 \\
\end{longtable}
\end{center}

\section{Digital Maturity and Technical Inefficiency}

Figure~\ref{fig:scatter_dma} presents a scatter plot of posterior mean
technical inefficiency against the digital maturity composite score for
all 111 trusts in the analytical sample. The Pearson correlation is
$r = -0.726$ ($p < 0.001$). The ordinary least squares slope is
$-5.1$ percentage points of inefficiency per unit increase in the
composite score (scale 1--5), indicating that a trust moving from the
sample minimum to the sample maximum digital maturity score (a range of
1.77 points) is associated with a reduction in estimated inefficiency of
approximately 9 percentage points. The dashed line shows the OLS fit and
is included for illustrative purposes; causal interpretation is not
warranted.

\begin{figure}[htbp]
\centering
\includegraphics[width=0.85\textwidth]{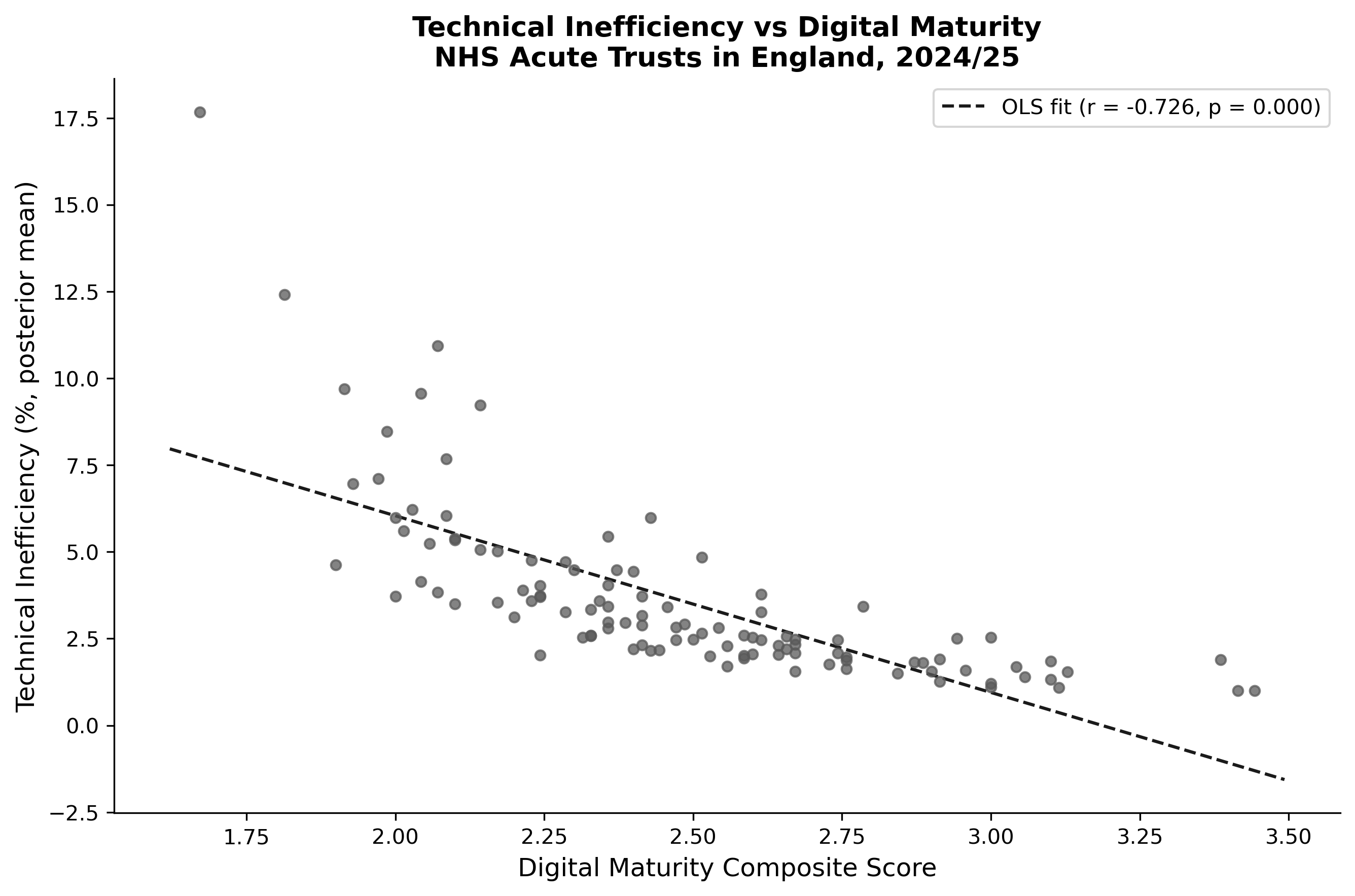}
\caption{Posterior mean technical inefficiency against digital maturity
composite score for 111 NHS acute non-specialist trusts in England,
2024/25. Inefficiency is expressed as a percentage of the production
frontier ($100 \times (1 - \text{TE}_i)$). The dashed line shows the
OLS fit ($r = -0.726$, $p < 0.001$). Efficiency scores from Model A
(four-input Cobb--Douglas, composite digital maturity specification).}
\label{fig:scatter_dma}
\end{figure}

\section{Capital and Digital Maturity: Correlation Analysis}

Table~\ref{tab:capital_corr} presents Pearson correlations between physical
capital, capital intensity, and digital maturity scores across the 111-trust
analytical sample. Capital intensity is defined as physical capital divided
by cost-weighted activity. The digital maturity composite is positively but
non-significantly correlated with physical capital stock ($r = 0.161$,
$p = 0.090$). One pillar -- Empower People -- reaches nominal significance
($r = 0.289$, $p = 0.002$) but does not survive Bonferroni correction for
the 16 tests conducted ($p < 0.006$ threshold). All correlations between
capital intensity and DMA scores are small and non-significant. These results
indicate that capital endowment is not a systematic correlate of digital
maturity in the analytical sample, though a weak association between capital
stock and the Empower People pillar warrants acknowledgement.

\begin{table}[htbp]
\centering
\caption{Correlations Between Physical Capital and Digital Maturity
         (111-Trust Analytical Sample)}
\label{tab:capital_corr}
\begin{threeparttable}
\begin{tabular}{lrrrr}
\toprule
& \multicolumn{2}{c}{Physical capital} &
  \multicolumn{2}{c}{Capital intensity} \\
\cmidrule(lr){2-3}\cmidrule(lr){4-5}
Variable & $r$ & $p$ & $r$ & $p$ \\
\midrule
Composite score      & $+$0.161 & 0.090 & $-$0.036 & 0.706 \\
Well Led             & $+$0.017 & 0.858 & $-$0.008 & 0.930 \\
Smart Foundations    & $+$0.111 & 0.245 & $+$0.006 & 0.946 \\
Safe Practice        & $+$0.185 & 0.052 & $+$0.049 & 0.609 \\
Support Workforce    & $-$0.018 & 0.850 & $-$0.182 & 0.056 \\
Empower People       & $+$0.289 & 0.002$^\dagger$ & $+$0.137 & 0.151 \\
Improve Care         & $+$0.107 & 0.266 & $-$0.014 & 0.882 \\
Healthy Populations  & $+$0.142 & 0.136 & $-$0.138 & 0.149 \\
\bottomrule
\end{tabular}
\begin{tablenotes}
\small
\item Capital intensity: physical capital divided by cost-weighted
activity. $^\dagger$Nominally significant at $p < 0.05$ but does not
survive Bonferroni correction for 16 simultaneous tests
($p < 0.006$ threshold). One significant result in 16 tests is
consistent with the expected false discovery rate under the null
hypothesis of no association.
\end{tablenotes}
\end{threeparttable}
\end{table}

\begin{figure}[htbp]
\centering
\includegraphics[width=\textwidth]{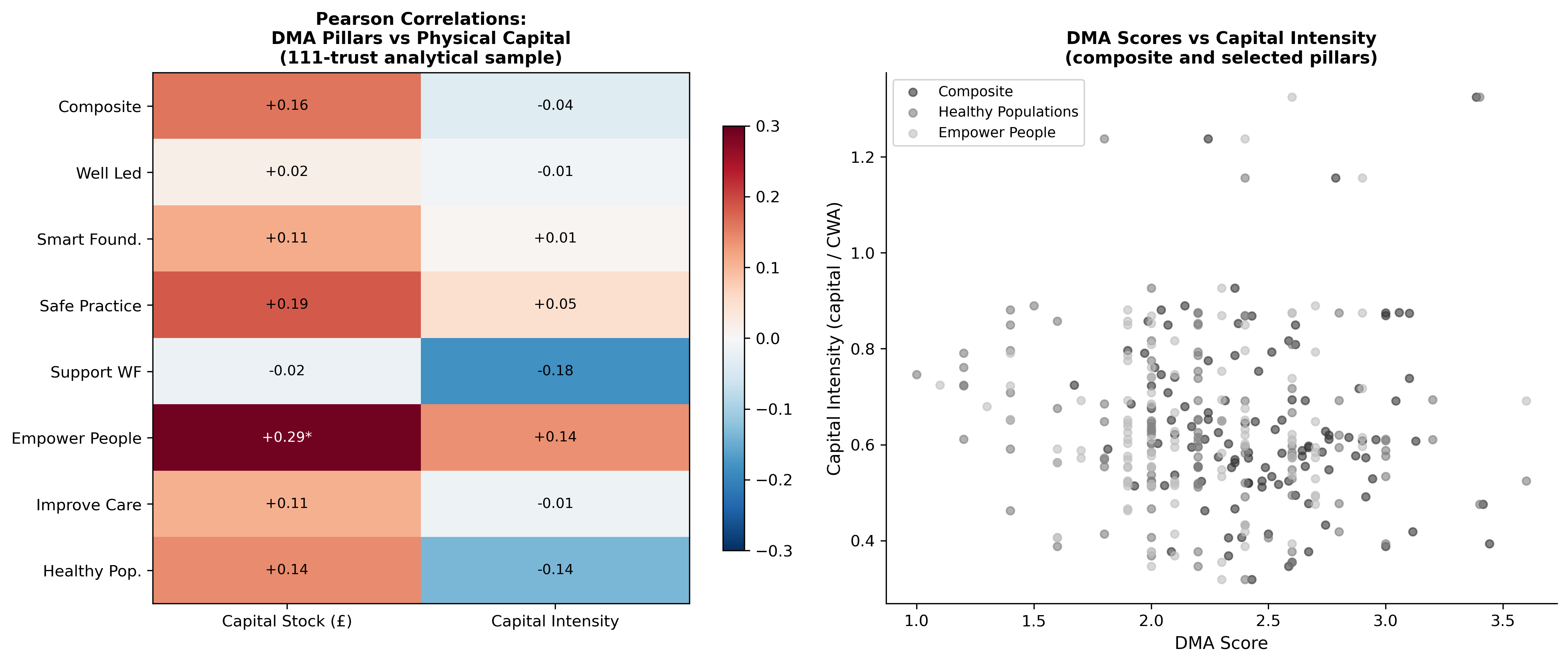}
\caption{Left: Pearson correlations between DMA pillar scores and physical
capital measures across the 111-trust analytical sample. Cells marked with
an asterisk reach nominal significance ($p < 0.05$); none survive Bonferroni
correction. Right: scatter plot of DMA scores against capital intensity
(physical capital divided by cost-weighted activity) for the composite score
and two selected pillars. The absence of a systematic relationship supports
the assumption that digital maturity is not a proxy for capital endowment.}
\label{fig:capital_corr}
\end{figure}

\section{Digital Maturity and Technical Inefficiency}

Figure~\ref{fig:scatter_dma} presents a scatter plot of posterior mean
technical inefficiency against the digital maturity composite score for
all 111 trusts in the analytical sample. The Pearson correlation is
$r = -0.726$ ($p < 0.001$). The ordinary least squares slope is
$-5.1$ percentage points of inefficiency per unit increase in the
composite score (scale 1--5), indicating that a trust moving from the
sample minimum to the sample maximum digital maturity score (a range of
1.77 points) is associated with a reduction in estimated inefficiency of
approximately 9 percentage points. The dashed line shows the OLS fit and
is included for illustrative purposes; causal interpretation is not
warranted.

\begin{figure}[htbp]
\centering
\includegraphics[width=0.85\textwidth]{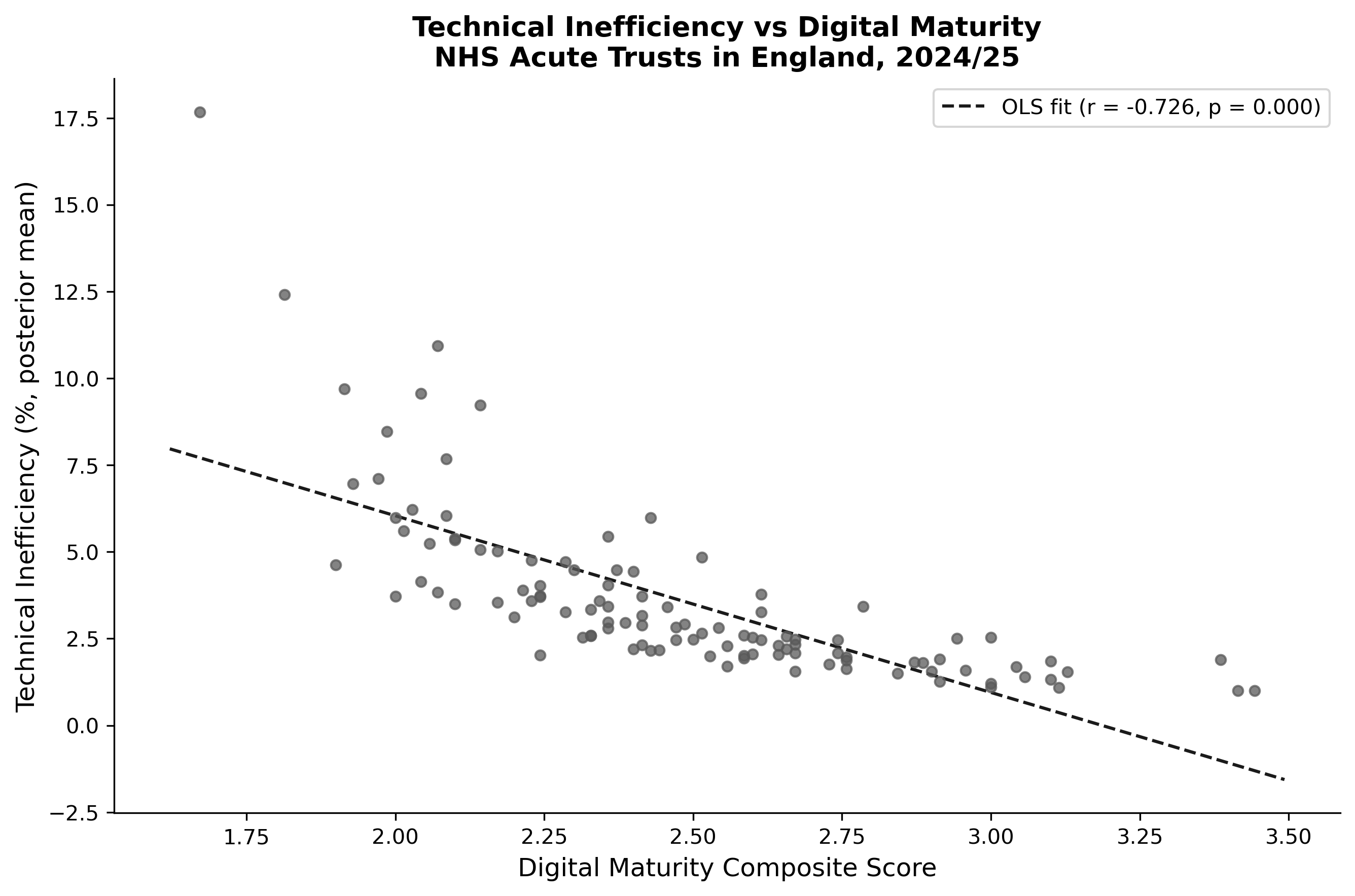}
\caption{Posterior mean technical inefficiency against digital maturity
composite score for 111 NHS acute non-specialist trusts in England,
2024/25. Inefficiency is expressed as a percentage of the production
frontier ($100 \times (1 - \text{TE}_i)$). The dashed line shows the
OLS fit ($r = -0.726$, $p < 0.001$). Efficiency scores from Model A
(four-input Cobb--Douglas, composite digital maturity specification).}
\label{fig:scatter_dma}
\end{figure}

\section{Package versions}

Analyses were conducted in Python 3.13.9 (Anaconda distribution, GCC 11.2,
Linux). Key packages: PyMC 6.0.1 \citep{abril2023pymc}, ArviZ 1.1.0,
PyTensor 3.0.3, NumPy 2.2.6, Pandas 2.3.3, SciPy 1.16.3, Matplotlib 3.10.6,
openpyxl 3.1.5, thefuzz 0.22.1, scikit-learn 1.6.1. Sampling was performed
using four CPU chains in parallel.

\newpage



\end{document}